# Title page

**Disruption of the mitochondrial network in a mouse model of Huntington's Disease visualized by in-tissue multiscale 3D electron microscopy**

**Authors:**

E. Martin-Solana[1], L. Casado-Zueras[2], T.E. Torres[3], G.F. Goya[2,4], M.R. Fernandez-Fernandez[5,*], J.J. Fernandez[5,*]

**Affiliations:**

[1] Dept Psychiatry, University of Pittsburgh, Pittsburgh, PA 15213, USA.

[2] Advanced Microscopy Laboratory, University of Zaragoza, Zaragoza, Spain.

[3] Institute of Nanoscience and Materials of Aragon (INMA). University of Zaragoza, 50018 Zaragoza, Spain.

[4] Dept. of Condensed Matter Physics, University of Zaragoza, Zaragoza, Spain.

[5] Spanish National Research Council (CSIC, CINN), Health Research Institute of Asturias (ISPA). 33011 Oviedo, Spain.

[*] **Corresponding authors:** MR.Fernandez@csic.es , JJ.Fernandez@csic.es

**Abstract:**

Huntington's disease (HD) is an inherited neurodegenerative disorder caused by an expanded CAG repeat in the coding sequence of huntingtin protein. Initially, it predominantly affects medium-sized spiny neurons (MSSNs) of the corpus striatum. No effective treatment is still available, thus urging the identification of potential therapeutic targets. While evidence of mitochondrial structural alterations in HD exists, previous studies mainly employed 2D approaches and were performed outside the strictly native brain context.

In this study, we adopted a novel multiscale approach to conduct a comprehensive 3D *in situ* structural analysis of mitochondrial disturbances in a mouse model of HD. We investigated MSSNs within brain tissue under optimal structural conditions utilizing state-of-the-art 3D imaging technologies, specifically FIB/SEM for the complete imaging of neuronal somas and Electron Tomography for detailed morphological examination, and image processing-based quantitative analysis.

Our findings suggest a disruption of mitochondrial morphology and dynamics towards fragmentation in HD. The network of interlaced, slim and long mitochondria observed in healthy conditions transforms into isolated, swollen and short entities, with internal cristae disorganization, cavities and abnormally large matrix granules.

This approach holds promise in opening new avenues for the *in situ* analysis of the subcellular disturbances and the identification of therapeutic targets in HD.

# 1. Introduction

Huntington's disease (HD) is an inherited neurodegenerative disorder caused by an expanded CAG repeat in the coding sequence of huntingtin protein (Htt). Symptoms typically emerge in middle age (35-45 years) and life expectancy post-onset is generally 15–20 years. Initially, the disease primarily affects the corpus striatum involving the selective neurodegeneration of medium-sized spiny neurons (MSSNs) (Bates et al., 2015) [1]. Clinical manifestations include motor, cognitive, and psychiatric symptoms. Despite extensive research since the discovery of its genetic cause, the precise pathophysiological mechanisms of HD remain poorly understood (Saudou and Humbert, 2016) [2]. Significantly, there is still no effective treatment for the disease (Kwon, 2021; Nowogrodzki, 2018) [3,4]. Consequently, identifying potential targets for therapeutic intervention in HD remains a top priority.

The expansion of a CAG triplet repeat in the sequence of the gene coding for Htt leads to an abnormally expanded polyglutamine (polyQ) tract at the N-terminus of the protein (The Huntington's Disease Collaborative Research Group, 1993) [5]. The aberrant polyQ expansion induces conformational changes in Htt that increase its propensity to form aggregates, hallmarks of the pathology. Htt is a ubiquitous, mainly cytoplasmic, protein and it has been related to the endoplasmic reticulum (ER), mitochondria and Golgi complex (Costa and Scorrano, 2012) [6]. Mutant Htt (mHtt), in either soluble or aggregate state, interferes with a wide spectrum of cellular functions: transcription, cell traffic, autophagy, metabolism, etc. (Bates et al., 2015) [1].

Mitochondria play a pivotal role in neurons as organelles responsible for meeting the high energy demands necessary to support their physiological functions. Evidence of mitochondrial dysfunction has been identified in HD (Dai et al., 2023; Neueder and Orth, 2020) [7,8]. However, the precise causes and nature of this dysfunction remain unknown, even with conflicting results among different systems used for investigation (Costa and Scorrano, 2012; Hering et al., 2017; Neueder and Orth, 2020; Oliveira, 2010; Pellman et al., 2015; Reddy, 2014) [6,7,9–12]. It is evident, nonetheless, that mitochondrial disturbances impact various aspects both structurally and functionally (Dai et al., 2023; Neueder and Orth, 2020) [7,8].

At the structural level, alterations in mitochondrial morphology and dynamics have been observed. Mitochondria are highly dynamic organelles that continuously undergo remodelling, fusion, fission, and migration. There is evidence that mitochondrial dynamics is disrupted in HD, with an imbalance between fusion and fission that results in excessive mitochondrial fragmentation driven by increased GTPase activity of Drp1, a protein implicated in fission (Cherubini et al., 2020; Costa et al., 2010; Dai et al., 2023; Guedes-Dias et al., 2016; Neueder et al., 2024; Reddy, 2014; Song et al., 2011) [8,10,13–17]. Ultimately, this may result in an abnormal distribution of mitochondria across the neuron domains. Morphologically, cristae disorganization and mitochondrial swelling have been reported, which impact their capacity to produce energy (Costa et al., 2010; Hering et al., 2017; Reddy, 2014; Song et al., 2011) [10,12–14]. Additionally, recent findings describe the enlargement of mitochondrial matrix granules (Wu et al., 2023) [18].

Electron microscopy (EM) stands as a classical technique for studying the cellular ultrastructure. Recent revolutionizing advances in three-dimensional (3D) imaging by EM and in sample preparation are enabling 3D ultrastructural studies of samples in their native context, preserved at close-to-physiological conditions and at a resolution of few nanometers. Electron Tomography (ET) and Focused Ion Beam Scanning Electron Microscopy (FIB/SEM) are two major 3DEM techniques that are allowing addressing fundamental questions in molecular and cell biology (Collinson et al., 2023; Fernandez and Martinez-Sanchez, 2022; Mahamid et al., 2016; Peddie et al., 2022) [19–22]. ET relies upon a transmission electron microscope (TEM) and provides 3D structural information with resolution around 2-4 nm from samples with limited thickness (250-500 nm). FIB/SEM overcomes this limitation by cyclically (i) milling a thin layer of the specimen using the FIB, followed by (ii) SEM imaging of the exposed surface. FIB/SEM thus collects information from large 3D volumes (tens of microns thick) at a resolution around 10 nm. These 3DEM techniques can be combined through multiscale integrative approaches so as to enable comprehensive ultrastructural studies.

Sample preparation constitutes a crucial step in EM. Cryofixation, involving the rapid freezing of the sample (in milliseconds) and maintaining it hydrated in vitreous ice, ensures optimal structural preservation, avoiding artefacts induced by traditional chemical fixatives (Korogod et al., 2015) [23]. Cryofixation of thick samples (up to 200 microns thick) is accomplished by high-pressure freezing (HPF) to prevent the ice crystal formation (Studer et al., 2014, 2008) [24,25]. While observing pristine HPF samples under cryogenic conditions would be ideal, it remains challenging for tissues, although it is increasingly feasible for cell cultures. Consequently, the standard protocol for tissues continues with freeze-substitution (FS) of the frozen water by an organic solvent and resin embedding (Fernandez-Fernandez et al., 2017) [26]. The HPF/FS tissue sample can then be (i) cut into thin sections (up to 200-500 nm) for observation with ET or (ii) directly visualized in the FIB/SEM at room temperature.

3DEM combined with sample cryofixation is providing new insights into cellular compartments and their functions (Brandt et al., 2017; Chen et al., 2023; Gemmer et al., 2023; Mahamid et al., 2016) [19,27–29] and is gaining traction in the exploration of neurodegenerative diseases (Leistner et al., 2023; Siegmund et al., 2018; Zhu et al., 2021) [30–32]. In HD in particular, these techniques are expanding our understanding of polyQ aggregates and subcellular alterations by working with *in vitro* samples or cultured cells (Bäuerlein et al., 2017; Galaz-Montoya et al., 2021; Wu et al., 2023; Zhao et al., 2023) [18,33–35]. It is important to note, however, that these studies are conducted outside the strictly native context, as they do not deal with intact tissue samples.

In this study, our objective is to investigate the structural perturbations of mitochondria under pathological conditions in HD within their native tissue environment. To this end, we conduct *in situ* 3D structural analysis of mouse brain tissue samples under optimal structural preservation conditions using a multiscale combination of advanced 3DEM techniques.

## 2. Results

**FIB/SEM tomography reveals alterations in the mitochondrial network and morphology in HD**

HD involves the selective degeneration of striatal MSSNs from the onset of the disease (Bates et al., 2015) [1]. So, we aimed at examination of the organization and morphology of mitochondria in these neurons within their native brain tissue context to identify and characterize their disturbances in HD. To study such a complex and large structure as the mitochondrial network in a comprehensive manner and in 3D, we used FIB/SEM tomography. This imaging technology allows us to image large samples (in the range of tens or hundreds of microns) at a resolution of few nanometers.

Brain tissue samples from a 10-months old HD animal model (heterozygous zQ175) (Menalled et al., 2012) [36] and a corresponding littermate wild-type (WT) control were prepared with HPF/FS, as described in Materials and Methods. This preparation protocol ensures optimal structural preservation of tissue samples. The tissue blocks were examined by FIB/SEM tomography and stacks were acquired from cells compatible with striatal MSSNs. A total of 8 and 5 MSSNs from the HD animal model and the control, respectively, were imaged. The acquired stacks represented volumes of a thickness in the range 5-30 microns.

Figure 1 presents representative 2D slices of three volumes, one MSSN from the WT animal (A) and two MSSNs from the HD model (B,C). In these slices, mitochondria are discernible as light grey masses within the darker spotty cytoplasm. Selected areas with representative mitochondria and their 3D context are also depicted in Figure 1 (D,E,F), illustrating different scenarios such as interlaced mitochondria (D), potential fission process (E) or an isolated dysmorphic entity (F). Figure 2 showcases a gallery of typical mitochondria collected from the slices of all the 13 volumes acquired. Figures 1 and 2 present consistent phenotypical mitochondrial features in the HD model and control. Mitochondria in the WT animal appear as relatively slim rods that, depending on the orientation in the 3D volume, may be observed as circular, elliptical or long shapes in the slices. Furthermore, they exhibit a nearly homogeneous inner density, except for the bright mitochondrial matrix granules, owing to their tightly packed cristae that are barely visible individually. In contrast, mitochondria in the HD model look as swollen, rough, irregular, and often grotesque shapes. Their inner density is not homogeneous at all. Instead, the cristae appear separated and can even be discerned by eye. Moreover, a distinctive feature is the presence of holes in the matrix, often with fuzzy material (Figure 2, arrows). Despite the resolution limitations, the matrix granules seem to be somewhat more apparent than in the WT animal.

Figure 3 depicts 3D views of the three representative MSSNs in Figure 1, showcasing segmented mitochondria and delineated plasma and nuclear membranes. The membranes are presented with transparency to ensure visibility of mitochondria at any side of the nucleus. This visualization of the entire cytoplasmic area that was imaged with the FIB/SEM microscope allows a more complete interpretation of the spatial distribution of mitochondria or their interrelationships. The thickness (size along the Z dimension) of those volumes was 25, 13 and 10 microns, respectively. In the WT animal, mitochondria appear as long and slim rods intricately interlaced and distributed throughout the cytoplasm, forming a complex mitochondrial network. In contrast, in the HD model, most mitochondria appear as isolated individual entities with irregular, rough and bumpy shapes and relatively short extensions, thus giving the impression of a disrupted mitochondrial network. While these alterations can somewhat recognizable in the 2D slices presented in Figures 1 and 2, the full interpretation of these changes is only achievable through the 3D visualization of a significant area of the neuronal soma as in Figure 3.

**Quantification of FIB/SEM data confirms the mitochondrial alterations in HD**

To conduct an objective quantification of the alterations, we devised a workflow that started with the automated segmentation of mitochondria in the volumes by means of an artificial-intelligence-based approach (see Materials and Methods, and also Figure 3 for the segmentation result). The binary volumes with the mitochondria segmented were then fed to MitoGraph software (Harwig et al., 2018; Rafelski et al., 2012) [37,38]. This program estimates the skeleton of each individual mitochondrion, given as edges (i.e. mitochondrial segments or branches) and nodes (i.e. terminal ends and branching points), and provides both length and local width of the individual edges (Figure 4(inset)). Finally, all the data from MitoGraph were compiled to provide measurements for each individual mitochondrion, namely number of edges (i.e. segments or branches), volume, length (sum of the length of its edges) and width (average of the width along the entire skeleton) (Figure 4(inset)).

To perform the quantification, we selected three representative FIB/SEM volumes, one MSSN from the control and two MSSNs from the HD model (Figure 3), with a significant cytoplasmic volume of the neuronal soma, 796, 588 and 385 µm$^3$, respectively. The automated segmentation procedure identified and segmented a total of 260, 140 and 119 mitochondria from the three volumes, respectively, which were then processed with MitoGraph. Subsequently, MitoGraph data were post-processed to obtain measurements for all individual mitochondria. For the statistical analysis, they were grouped into two categories: WT (260 mitochondria) and HD (259).

Figure 4 presents the quantification results. While there are no significant differences in the total volume per mitochondrion between the phenotypes, there is a trend towards higher volume in HD. However, the other measurements indeed reveal statistically significant differences. Remarkably, mitochondria in the HD model are shorter and exhibit a lower number of edges, mostly one, whereas in the WT counterpart they are notably longer and consist of a higher number of segments or branches, up to 10. This observation strongly supports the idea that an intricate network composed of mitochondria with many and relatively long branches is disrupted under pathological conditions in HD, transforming mitochondria into short, monolithic individual entities. This is indicative of a scenario of mitochondrial fragmentation.

Moreover, Figure 4 also shows that mitochondria under healthy conditions are relatively thin, mostly with a width around 0.17 µm. In contrast, in the HD model they are significantly thicker, with a broad width range (up to 0.44 µm) that accounts for irregular and bumpy shapes. Visualization of the length and width measurements of individual mitochondria in a scatterplot, as presented in Figure 4 (bottom right), clearly enables identification of the two phenotypes: short and thick in HD whereas long and slim in WT. These results explain why no significant differences in the total volume per mitochondrion are found, as described previously.

In summary, these measurements and the results in Figure 4 faithfully reflect in objective and quantitative terms the alterations of the network and morphology observed in Figures 1-3. They indicate disruptions in mitochondrial dynamics, with a tendency towards fragmentation, and disturbances in morphology, characterized by shorter and thicker (swollen) mitochondria under pathological conditions in HD.

**Electron tomography enables zoomed-in analysis of the mitochondrial disturbances and quantification of the altered matrix granules in HD.**

Although FIB/SEM tomography was valuable in visualizing the mitochondrial network and morphology, as described in previous sections, its resolution was still limited for studying fine details of the mitochondrial matrix, particularly granules. Therefore, we switched to ET as it enables identification and characterization of alterations in matrix granules thanks to the higher resolution power of this imaging technique.

For the ET experiments, we kept focused on striatal MSSNs from brain tissue samples and had available a 9-months old HD animal model (homozygous zQ175) and a corresponding WT control. The samples were prepared with HPF/FS, thereby ensuring optimal structural preservation as already described, and cut into ultrathin 250-nm-thick sections suitable for TEM and ET. A total of 120 sections from cells compatible with striatal MSSNs of the HD model and control were observed by TEM, showing consistent structural patterns in accordance with the observations by FIB/SEM already described. Subsequently, representative areas containing mitochondria from the HD model and the control were selected for examination through ET.

Figure 5 presents representative slices of the tomograms of the MSSNs from the WT animal (A) and the HD model (B) as well as their 3D visualization (C). Note that the contrast in TEM and ET is the opposite of that in FIB/SEM, hence electron-dense material appears as darker features in the images. The mitochondria in Figure 5 exhibit features consistent with the previous FIB/SEM results shown in Figures 1 and 2. In WT, a bundle of mitochondria was imaged, all of them being slim and rod-shaped and having relatively homogeneous density, except for the granules. The cristae can still be glimpsed, displaying a compact stacked organization. However, the mitochondrion from the HD model appears aberrantly swollen with a disrupted matrix where the cristae are abnormally disorganized and significant areas are devoid of material or have fuzzy content (Figure 5, arrow).

ET allowed identification of matrix granules and thus a fair comparison between the phenotypes. Visually, the dense granules exhibit granulated textures, consisting of smaller structural units (Figure 5, insets). Furthermore, Figure 5 already shows that the granules in the HD model generally appear larger than in the control. To carry out a quantitative analysis about their size, the individual granules were segmented in objective manner (as described in Materials and Methods), yielding a total of 66 and 37 granules in the WT and HD model, respectively, and their volumes were then measured. Figure 5 (D) presents the quantification results that confirm significantly enlarged granules in HD. In the WT animal there is a narrow distribution of granule volumes with a median around 46000 $nm^3$ (equivalent to a sphere with diameter 44 nm) whereas in the HD model the distribution is much wider and with a median around 84000 $nm^3$ (equivalent to a sphere with diameter 54 nm).

## 3. Discussion and conclusion

There is evidence of mitochondrial dysfunction in HD, with implications at the functional and structural levels (Dai et al., 2023) [8]. However, certain conclusions remain controversial, likely caused by the inherent differences between model systems or experimental approaches (Neueder and Orth, 2020)[7]. In this study, we employed a novel multiscale approach for a comprehensive 3D *in situ* structural study of the mitochondrial disturbances in a mouse model of HD. We analyzed MSSNs from brain tissue samples prepared with cryo-fixation-based methods to ensure structural preservation at close-to-native conditions. By combining various state-of-the-art 3D imaging technologies, we examined mitochondria over significantly large areas of the neuronal soma (FIB/SEM) and their inner details with sufficient resolution (ET). Finally, computational image processing facilitated quantification of the disturbances.

Our findings suggest a disruption of mitochondrial dynamics in HD, leading to fragmentation. This results in isolated, short, swollen and aberrantly shaped mitochondria dispersed throughout the cytoplasm, where the intricate network composed of interlaced, slim, long and branched mitochondria found in healthy conditions no longer exists. Moreover, upon closer examination, swollen mitochondria in HD exhibit disorganized cristae, internal hollow areas with fuzzy contents and abnormally large matrix granules. In contrast, in the control they appear dense with tightly stacked cristae and granules of moderate size.

One strength of our approach is that it considers large areas of the neuronal soma and works in 3D, allowing for a more precise analysis of mitochondrial dynamics. Working in 2D, hence from only partial views of mitochondria obtained from ultrathin sections of cells, is prone to misleading results concerning mitochondrial dynamics, as it would be the case if conclusions on fragmentation were to be taken from Figure 2. In our approach, image processing has also been important for quantification. However, to quantitatively reflect the visual results (Figure 3), we had to employ elaborate metrics such as number of edges, length and width of mitochondria (Figure 4). In this regard, the total volume per mitochondrion, which is the simplest measurement, proved to be a misleading metric that did not adequately reflect the evident alterations observed in Figure 3. It showed no differences between control and HD (Figure 4) simply due to the transformation of mitochondria from slim and long to swollen and short in HD. Regardless, our results of mitochondrial fragmentation are consistent with previous studies on other HD models or human samples, primarily based on 2D microscopy or molecular approaches (e.g. Cherubini et al., 2020; Costa et al., 2010; Neueder et al., 2024; Song et al., 2011) [13,14,16,17]. Conflicting results also exist (Hering et al., 2017) [12], possibly attributable to differences in HD models, experimental strategies or the reasons just mentioned in this paragraph.

Regarding morphology, the mitochondrial swelling and cristae disorganization observed in MSSNs (Figures 2 and 5) align with all morphological reports thus far (e.g. Costa et al., 2010; Hering et al., 2017; Miguez et al., 2023; Reddy, 2014) [10,12,13,39]. Remarkably, we noticed hollow areas in the matrix with fuzzy content. They resemble mitochondrial vacuolization scenarios highlighted in previous works (Hering et al., 2017; Miguez et al., 2023) [12,39]. They are also compatible with the mitochondrial pockets that emerge under perturbed mitochondrial dynamics to encompass the aberrant accumulation of mitochondrial RNA granules (MRG, fluid condensates that comprise essential components of the mitochondrial post-transcriptional pathway and mitoribosome biogenesis) (Rey et al., 2020) [40].

The texture and volume of the matrix granules as well as the disturbances observed in intact brain tissue (Figure 5) are consistent with recent findings in mouse neuronal cultures by cryo-ET (Wu et al., 2023) [18]. These granules, identified as calcium phosphate deposits, are related to the role of mitochondria in subcellular calcium homeostasis. They absorb excess cytoplasmic calcium and store it in the form of granules (Panov et al., 2002; Pivovarova and Andrews, 2010) [41,42], whose larger size needs accommodation in the matrix through remodelling cristae. An excessive calcium influx can trigger mitochondrial swelling, depolarization and eventual collapse (Pivovarova and Andrews, 2010; Strubbe-Rivera et al., 2021) [42,43], potentially contributing to the observed aberrant morphology. The composition of these calcium phosphate granules has been determined by elemental analysis (Wolf et al., 2017) [44] and studies on mitochondrial calcium uptake capacity with cryo-fixed samples (Kristian et al., 2007; Strubbe-Rivera et al., 2021) [43,45]. Therefore, the abnormal size of the granules in the HD model may be linked to the excess of basal intracellular calcium (Czeredys, 2020; Quintanilla et al., 2013; Raymond, 2017) [46–48] or the controversially reported lower calcium uptake capacity (Panov et al., 2002; Pellman et al., 2015) [11,41] associated to this disease. Of note, visualization and quantification these granules are only possible with cryo-fixed samples (Kristian et al., 2007; Strubbe-Rivera et al., 2021; Wolf et al., 2017; Wu et al., 2023) [18,43–45] since chemical fixation methods extract them (Pivovarova et al., 2004) [49].

Recently, new proteomics data have revealed an enrichment of proteins involved in RNA processing in HD patient iPSC-derived cultured neurons, leading to suggest that the matrix granules are MRGs (Wu et al., 2023) [18]. Therefore, there exists the possibility that both, calcium phosphate and components of MRGs, coexist as constituents of the observed electron-dense granules. Alternatively, the abnormally accumulated MRGs might be located within the mitochondrial pockets (Rey et al., 2020) [40] often observed in our FIB/SEM volumes, as described above.

In conclusion, our approach combining tissue cryofixation, multiscale 3D electron microscopy and image processing has enabled for the first time direct visualization, holistic analysis and quantification of the mitochondrial disruptions in HD within the native brain context. Our results provide a satisfactory explanation for the assumptions in HD research derived from partial or 2D approaches and might serve to reconcile some of the conflicting views, mitochondrial fragmentation in particular. Our innovative approach holds promise in opening new avenues for the *in situ* analysis of the disturbances of subcellular compartments and the identification of therapeutic targets in HD and in other neurodegenerative diseases, as many of them share similar hallmarks.

## Materials and Methods

### Animals

A stable colony of the zQ175 mice (Menalled et al., 2012) [36] was established through founders donated by the Cure Huntington's Disease Initiative (CHDI) and sourced from Jackson Laboratory Inc. The zQ175 line is a knock-in model bred on a C57BL/6J background featuring an endogenous murine *HTT* gene with a chimeric human/mouse exon 1 containing approximately 190 CAG repeats (B6.12951-Htt<tm1Mfc<190JChdi). Heterozygous and homozygous mice and wild-type (WT) control counterparts were bred to maintain a stable colony within the animal facility of the Centro Nacional de Biotecnologia (CSIC). They were provided with food and water *ad libitum*. All experiments complied with Spanish and European legislation and were in accordance with the ethical guidelines established by the Spanish National Research Council (CSIC) ethics committee concerning animal experimentation.

### Sample preparation based on HPF/FS

Brain tissue samples were prepared for ET and FIB/SEM imaging following our established protocols designed to ensure optimal structural preservation. These procedures are primarily based on high-pressure freezing and freeze-substitution (HPF/FS), as previously described (Fernandez-Fernandez et al., 2017) [26]. In short, mouse brains were dissected immediately post-mortem and 200-µm-thick sagittal slices were cut using a tissue slicer (Stoelting, Co.). Striatal samples were promptly extracted, placed onto a flat specimen carrier, and then subjected to high-pressure freezing within a Leica EMPACT2 device. The samples were further processed with freeze-substitution of frozen water to methanol and were subsequently embedded in Lowicryl resin HM20 with a Leica AFS2 EM FSP system.

For visualization in the TEM and for ET, sections (250 nm thick) were obtained from the resin-embedded samples using a Leica Ultracut EM-UC6 ultramicrotome, and placed on Quantifoil S7/2 grids.

### FIB/SEM imaging

FIB/SEM imaging was done on a FEI/ThermoFisher Scientific Helios NanoLab Dual-Beam 650 at the LMA node of the Spanish ICTS ELECMI. Regions of interest were identified by visual inspection of the sample surface with the SEM. Areas with cells compatible with striatal MSSNs were selected based on morphological criteria (Matamales et al., 2009) [50] and FIB/SEM stacks were then acquired. Prior to imaging, the areas were coated with a protective layer of carbon. The milling was performed using a slice thickness of 15 to 25 nm. Stacks of hundreds of images representing volumes of thickness 5 to 30 microns were acquired, with a pixel size at the specimen level in the range of 8 to 11 nm.

**TEM and ET imaging**

A conventional JEOL JEM-1011 transmission electron microscope (100 kV) was used to screen the 250-nm-thick sections, check the integrity of the tissue samples, and select areas of interest. Cells compatible with striatal MSSNs were selected based on morphological criteria (Matamales et al., 2009) [50] for subsequent ET and analysis. The magnification was set to 10K and 30K for identification of neurons and mitochondria, respectively.

Tomographic data were acquired by taking series of images from the sections while tilting them within a range of ±60° at 1° interval around a single tilt-axis. The tilt-series were acquired using a Thermo Fisher Scientific/FEI Tecnai G2 (200 kV) equipped with a CCD camera. The pixel size at the specimen level was 0.79 nm. For processing, visualization and analysis, the images were rescaled with a binning factor of 4. Prior to ET, grids were incubated in a solution of 10-nm diameter colloidal gold (EM.BSA 10, Electron Microscopy Sciences, Hatfield, PA, USA) to facilitate subsequent image alignment.

**Image processing of FIB/SEM stacks**

Acquired stacks were first processed with contrast enhancement and noise reduction (Fernandez et al., 2020; González-Ruiz et al., 2023) [51,52]. The resulting stacks were then subjected to alignment with IMOD (Kremer et al, 1996) [53]. For 3D visualization and analysis, the stacks were rescaled to isotropic voxel size (González-Ruiz et al., 2022) [54]. Delineation of plasma and nuclear membranes was done manually with IMOD tools to define the cytoplasmic area of the neuron. Semantic segmentation of mitochondria in the cytoplasm was performed with automated deep-learning procedures. To this end, a 2.5D U-net neural network was implemented to operate on 2D slices plus the immediate neighbour slices to predict the location of mitochondria using Dragonfly software (Comet Technologies Canada Inc.) (Makovetsky et al., 2018) [55]. The U-net had a depth level of 5 and worked on patches of 64x64. The training was volume-specific and was conducted in two steps. For the first step, training data was objectively produced by a computational procedure consisting of edge-preserving filtering of the FIB/SEM stacks with anisotropic non-linear diffusion (Fernández and Li, 2003; Moreno et al., 2018) [56,57] followed by thresholding on density and on size of connected components (Martinez-Sanchez et al., 2014, 2011) [58,59]. This procedure enabled preliminary segmentation mitochondria. The U-net was then trained using a subset of 100 slices of these stacks, with 2x data augmentation, using a batch size of 32 and a maximum number of epochs of 25 with early stopping criterion on a 20% slice subset that acted as a validation subset. After that first training step, the labels predicted for the training subset were manually revised to produce a new training data. The U-net network was then subjected to a second training step using the new data, continuing from the previous state of the network and using the same training hyperparameters. The final trained network was then applied to the whole stack to derive, after some manual revision, the definite segmented mitochondria.

Quantitative analysis of the segmented mitochondria was then carried out with MitoGraph (Harwig et al., 2018; Rafelski et al., 2012) [37,38] and in-house programs. The binary, segmented tomograms were processed with MitoGraph to decompose the individual mitochondria into their body and their skeleton comprising edges (mitochondrial segments or branches) and nodes (either mitochondrial ends or branching points) and to obtain measurements of the length of edges and the local width (distance from the edge points to the mitochondrial surface). An in-house program was developed to process these data and to provide measurements for individual mitochondria (number of edges, volume, length as the sum of the length of their edges, and width as the average of the width along the entire skeleton) for statistical analysis and for visualization with IMOD.

**Image processing of ET data**

Alignment of the tilt-series and 3D reconstruction of the tomograms were conducted using IMOD software (Kremer et al, 1996) [53] and Tomo3D (Agulleiro and Fernandez 2015, 2011) [60,61] applying standard protocols (Fernandez, 2012) [62]. Alignment was based on the colloidal gold beads as fiducial markers using IMOD. Tomographic reconstruction relied on weighted back-projection (WBP) using a filter that simulates an iterative reconstruction method (SIRT).

Delineation of mitochondrial outer membranes in the tomograms was done manually with IMOD tools. Automated semantic segmentation of mitochondrial matrix granules was done with edge-preserving filtering of the tomograms with anisotropic non-linear diffusion (Fernández and Li, 2003; Moreno et al., 2018) [56,57] followed by density thresholding. This procedure was enough to segment the granules owing to their significantly different density in comparison with the rest of the mitochondrial matrix.

**Statistical analysis and plotting**

Statistical analyses were performed with Python using the Pingouin package (Vallat, 2018) [63]. The comparisons were carried out based on the Mann-Whitney test as the Shapiro-Wilk tests indicated that all data were non-normally distributed. Plots were generated with the Seaborn package (Waskom, 2021) [64].

## Acknowledgements

We thank Eber Martínez and all staff at the Animal Facility of the Centro Nacional de Biotecnología (CNB-CSIC) for their work with our mice over the years. We also thank the Electron Microscopy Facility at the CNB-CSIC for the HPF and FS experiments and the CryoEM Facility for acquisition of ET data. We are also thankful to the LMA node of the Spanish ICTS ELECMI for providing us with access to the FIB/SEM microscope through grant ELC30-2023. This work was supported through grant A-4206 from Cure Huntington's Disease Initiative (CHDI) Foundation, a grant from the Huntinton's Disease Society of America, grant CIVP18A3892 from Fundación Ramón Areces and through grants SAF2017-84565-R, TED2021-132020B-I00 and PID2022-139071NB-I00 funded by MCIN/AEI/10.13039/501100011033, "ERDF A way of making Europe" and by the "European Union NextGenerationEU/PRTR".

## Authors' contributions

(CRediT: https://onlinelibrary.wiley.com/doi/epdf/10.1002/leap.1210)

EMS: Investigation, Validation, Formal Analysis, Writing - Original draft

LCZ: Investigation, Resources

TET: Investigation, Resources

GFG: Investigation, Resources, Funding Acquisition.

MRFF: Conceptualization, Methodology, Investigation, Validation, Formal Analysis, Writing - Original draft, Funding Acquisition.

JJF: Conceptualization, Methodology, Software, Investigation, Validation, Formal Analysis, Writing - Original draft, Funding Acquisition.

All authors read and approved the final manuscript.

# Figure Legends

**Figure 1. FIB/SEM tomography of MSSNs.** (A,B,C) Two representative XY slices of one MSSN from the WT animal (A) and two MSSNs from the HD model (B,C) are shown. Green and cyan contours delineate the plasma and nuclear membranes, respectively. Mitochondria are identified as light grey masses inside the darker cytoplasm. Dashed boxes enclose selected cytoplasmic areas with representative mitochondria. Bar: 1 µm. **(**D,E,F) Magnified views of the dashed boxes in (A,B,C), respectively, are presented (left panels) along with 3D isosurface representations (right panels) of a volume of 2x2x2 µm$^3$ around those areas, thus showing the nearby context of those mitochondria.

**Figure 2. Mitochondria in slices of FIB/SEM volumes.** Gallery of characteristic mitochondria observed in 2D slices of all FIB/SEM volumes. Top: mitochondria from 5 MSSNs from the WT animal. Bottom: mitochondria from 8 MSSNs from the HD animal model. Arrows indicate some hollow areas with fuzzy content in the matrix. Bar: 1 µm.

**Figure 3. 3D visualization of the FIB/SEM volumes**. Three different views of the volumes from the MSSNs in Figure 1 (A,B,C) are presented in the top (WT), middle (HD) and bottom (HD) rows, respectively. The leftmost views show the volumes with their Z axis running through the depth, a 90º rotation around the horizontal axis results in the views at the central panels, and a subsequent 90º rotation around the vertical axis produces the rightmost views. Segmented mitochondria are depicted with isosurface representation in gold color. Plasma and nuclear membranes are displayed in 85% transparent green and 50% transparent cyan, respectively, allowing visualization of the mitochondria behind the nucleus. The missing wedge in the volume shown in the middle row (central panel) is caused by a technical drift while FIB/SEM acquisition. Bar: 1 µm.

**Figure 4. Quantification of mitochondrial alterations in FIB/SEM volumes. Measurements** (upper left inset). Each individual mitochondrion consists of its body and skeleton, where the skeleton comprises edges (mitochondrial segments or branches) and nodes (i.e. terminal ends and branching points). For each mitochondrion, the following measurements are obtained: number of edges, volume, length (sum of the length of its edges) and width (average of the local width -here shown with a colormap- along its edges). Illustrative examples of mitochondria from WT and HD animals are presented in semitransparent 3D isosurface representation with their skeleton overlaid. **Quantification plots.** Comparison of measurements based upon 260 and 259 mitochondria from MSSNs of the WT animal and HD model. Violin plots show the distribution of the mitochondrial measurements. A miniature boxplot is included inside the violin plots, with the box representing the interquartile range (between the first and third quartile), an additional quartile with the whiskers and the median with a white dot. The scatterplot at the bottom rightmost panel represents measurements (length and width) of all individual mitochondria. *p<0.05; **p<0.0001.

**Figure 5. Electron tomography of mitochondria from MSSNs and analysis of matrix granules.** (A,B) Tomograms of mitochondria from a WT animal (A) and a HD model (B). Three slices, separated by 22.12 nm, are shown for each tomogram. Note that the contrast is the opposite of that in FIB/SEM volumes (Figures 1 and 2), with electron-dense material appearing darker. The arrow points to a hollow area with fuzzy content. Dashed boxes indicate granules magnified in the insets. Bar: 0.5 µm. (C) 3D visualization of the matrix granules within mitochondria. Granules are presented with isosurface representation in gold color and mitochondrial membranes are delineated in different semitransparent colors. (D) The distribution of the granule volumes obtained from 66 and 37 granules of the WT and HD animals, respectively, are presented with violin plots (right). Similar to Figure 4, a miniature boxplot is included inside the violin plots and the median of the distributions denoted by a white dot. **$p<0.0001$.

## References

1. Bates, G. P. *et al.* Huntington disease. *Nat Rev Dis Primers* **1**, 15005 (2015).

2. Saudou, F. & Humbert, S. The Biology of Huntingtin. *Neuron* **89**, 910–926 (2016).

3. Nowogrodzki, A. Huntington's disease: 4 big questions. *Nature* **557**, S48–S48 (2018).

4. Kwon, D. Failure of genetic therapies for Huntington's devastates community. *Nature* **593**, 180–180 (2021).

5. Macdonald, M. A novel gene containing a trinucleotide repeat that is expanded and unstable on Huntington's disease chromosomes. *Cell* **72**, 971–983 (1993).

6. Costa, V. & Scorrano, L. Shaping the role of mitochondria in the pathogenesis of Huntington's disease: Mitochondrial and Huntington's disease. *The EMBO Journal* **31**, 1853–1864 (2012).

7. Neueder, A. & Orth, M. Mitochondrial biology and the identification of biomarkers of Huntington's disease. *Neurodegenerative Disease Management* **10**, 243–255 (2020).

8. Dai, Y. *et al.* A comprehensive perspective of Huntington's disease and mitochondrial dysfunction. *Mitochondrion* **70**, 8–19 (2023).

9. Oliveira, J. M. A. Nature and cause of mitochondrial dysfunction in Huntington's disease: focusing on huntingtin and the striatum. *Journal of Neurochemistry* **114**, 1–12 (2010).

10. Reddy, P. H. Increased mitochondrial fission and neuronal dysfunction in Huntington's disease: implications for molecular inhibitors of excessive mitochondrial fission. *Drug Discovery Today* **19**, 951–955 (2014).

11. Pellman, J. J., Hamilton, J., Brustovetsky, T. & Brustovetsky, N. $Ca^{2+}$ handling in isolated brain mitochondria and cultured neurons derived from the YAC 128 mouse model of Huntington's disease. *Journal of Neurochemistry* **134**, 652–667 (2015).

12. Hering, T. *et al.* Mitochondrial cristae remodelling is associated with disrupted OPA1 oligomerisation in the Huntington's disease R6/2 fragment model. *Experimental Neurology* **288**, 167–175 (2017).

13. Costa, V. *et al.* Mitochondrial fission and cristae disruption increase the response of cell models of Huntington's disease to apoptotic stimuli. *EMBO Mol Med* **2**, 490–503 (2010).

14. Song, W. *et al.* Mutant huntingtin binds the mitochondrial fission GTPase dynamin-related protein-1 and increases its enzymatic activity. *Nat Med* **17**, 377–382 (2011).

15. Guedes-Dias, P. *et al.* Mitochondrial dynamics and quality control in Huntington's disease. *Neurobiology of Disease* **90**, 51–57 (2016).

16. Cherubini, M., Lopez-Molina, L. & Gines, S. Mitochondrial fission in Huntington's disease mouse striatum disrupts ER-mitochondria contacts leading to disturbances in Ca2+ efflux and Reactive Oxygen Species (ROS) homeostasis. *Neurobiology of Disease* **136**, 104741 (2020).

17. Neueder, A. *et al.* Huntington disease affects mitochondrial network dynamics predisposing to pathogenic mtDNA mutations. *Brain* awae007 (2024) doi:10.1093/brain/awae007.

18. Wu, G.-H. *et al.* CryoET reveals organelle phenotypes in huntington disease patient iPSC-derived and mouse primary neurons. *Nat Commun* **14**, 692 (2023).

19. Mahamid, J. *et al.* Visualizing the molecular sociology at the HeLa cell nuclear periphery. *Science* **351**, 969–972 (2016).

20. Peddie, C. J. *et al.* Volume electron microscopy. *Nat Rev Methods Primers* **2**, 51 (2022).

21. Fernandez, J.-J. & Martinez-Sanchez, A. Computational methods for three-dimensional electron microscopy (3DEM). *Computer Methods and Programs in Biomedicine* **225**, 107039 (2022).

22. Collinson, L. M. *et al.* Volume EM: a quiet revolution takes shape. *Nat Methods* **20**, 777–782 (2023).

23. Korogod, N., Petersen, C. C. & Knott, G. W. Ultrastructural analysis of adult mouse neocortex comparing aldehyde perfusion with cryo fixation. *eLife* **4**, e05793 (2015).

24. Studer, D. *et al.* Capture of activity-induced ultrastructural changes at synapses by high-pressure freezing of brain tissue. *Nat Protoc* **9**, 1480–1495 (2014).

25. Studer, D., Humbel, B. M. & Chiquet, M. Electron microscopy of high pressure frozen samples: bridging the gap between cellular ultrastructure and atomic resolution. *Histochem Cell Biol* **130**, 877–889 (2008).

26. Fernandez-Fernandez, M. R. *et al.* 3D electron tomography of brain tissue unveils distinct Golgi structures that sequester cytoplasmic contents in neurons. *Journal of Cell Science* **130**, 83–89 (2017).

27. Brandt, T. *et al.* Changes of mitochondrial ultrastructure and function during ageing in mice and Drosophila. *eLife* **6**, e24662 (2017).

28. Chen, Z. *et al.* De novo protein identification in mammalian sperm using in situ cryoelectron tomography and AlphaFold2 docking. *Cell* S0092867423010383 (2023) doi:10.1016/j.cell.2023.09.017.

29. Gemmer, M. *et al.* Visualization of translation and protein biogenesis at the ER membrane. *Nature* **614**, 160–167 (2023).

30. Leistner, C. *et al.* The in-tissue molecular architecture of β-amyloid pathology in the mammalian brain. *Nat Commun* **14**, 2833 (2023).

31. Siegmund, S. E. *et al.* Three-Dimensional Analysis of Mitochondrial Crista Ultrastructure in a Patient with Leigh Syndrome by In Situ Cryoelectron Tomography. *iScience* **6**, 83–91 (2018).

32. Zhu, Y. *et al.* Serial cryoFIB/SEM Reveals Cytoarchitectural Disruptions in Leigh Syndrome Patient Cells. *Structure* **29**, 82-87.e3 (2021).

33. Bäuerlein, F. J. B. *et al.* In Situ Architecture and Cellular Interactions of PolyQ Inclusions. *Cell* **171**, 179-187.e10 (2017).

34. Galaz-Montoya, J. G., Shahmoradian, S. H., Shen, K., Frydman, J. & Chiu, W. Cryo-electron tomography provides topological insights into mutant huntingtin exon 1 and polyQ aggregates. *Commun Biol* **4**, 849 (2021).

35. Zhao, D. Y. *et al. Autophagy Preferentially Degrades Non-Fibrillar polyQ Aggregates.* http://biorxiv.org/lookup/doi/10.1101/2023.08.08.552291 (2023) doi:10.1101/2023.08.08.552291.

36. Menalled, L. B. *et al.* Comprehensive Behavioral and Molecular Characterization of a New Knock-In Mouse Model of Huntington's Disease: zQ175. *PLoS ONE* **7**, e49838 (2012).

37. Harwig, M. C. *et al.* Methods for imaging mammalian mitochondrial morphology: A prospective on MitoGraph. *Analytical Biochemistry* **552**, 81–99 (2018).

38. Rafelski, S. M. *et al.* Mitochondrial Network Size Scaling in Budding Yeast. *Science* **338**, 822–824 (2012).

39. Miguez, A. *et al.* Soluble mutant huntingtin drives early human pathogenesis in Huntington's disease. *Cell. Mol. Life Sci.* **80**, 238 (2023).

40. Rey, T. *et al.* Mitochondrial RNA granules are fluid condensates positioned by membrane dynamics. *Nat Cell Biol* **22**, 1180–1186 (2020).

41. Panov, A. V. *et al.* Early mitochondrial calcium defects in Huntington's disease are a direct effect of polyglutamines. *Nat Neurosci* **5**, 731–736 (2002).

42. Pivovarova, N. B. & Andrews, S. B. Calcium-dependent mitochondrial function and dysfunction in neurons. *The FEBS Journal* **277**, 3622–3636 (2010).

43. Strubbe-Rivera, J. O. *et al.* The mitochondrial permeability transition phenomenon elucidated by cryo-EM reveals the genuine impact of calcium overload on mitochondrial structure and function. *Sci Rep* **11**, 1037 (2021).

44. Wolf, S. G. *et al.* 3D visualization of mitochondrial solid-phase calcium stores in whole cells. *eLife* **6**, e29929 (2017).

45. Kristian, T., Pivovarova, N. B., Fiskum, G. & Andrews, S. B. Calcium-induced precipitate formation in brain mitochondria: composition, calcium capacity, and retention. *Journal of Neurochemistry* **102**, 1346–1356 (2007).

46. Czeredys, M. Dysregulation of Neuronal Calcium Signaling via Store-Operated Channels in Huntington's Disease. *Front. Cell Dev. Biol.* **8**, 611735 (2020).

47. Quintanilla, R. A., Jin, Y. N., Von Bernhardi, R. & Johnson, G. V. Mitochondrial permeability transition pore induces mitochondria injury in Huntington disease. *Mol Neurodegeneration* **8**, 45 (2013).

48. Raymond, L. A. Striatal synaptic dysfunction and altered calcium regulation in Huntington disease. *Biochemical and Biophysical Research Communications* **483**, 1051–1062 (2017).

49. Pivovarova, N. B. *et al.* Excitotoxic Calcium Overload in a Subpopulation of Mitochondria Triggers Delayed Death in Hippocampal Neurons. *J. Neurosci.* **24**, 5611–5622 (2004).

50. Matamales, M. *et al.* Striatal Medium-Sized Spiny Neurons: Identification by Nuclear Staining and Study of Neuronal Subpopulations in BAC Transgenic Mice. *PLOS ONE* **4**, e4770 (2009).

51. Fernandez, J.-J., Torres, T. E., Martin-Solana, E., Goya, G. F. & Fernandez-Fernandez, M.-R. PolishEM: image enhancement in FIB–SEM. *Bioinformatics* **36**, 3947–3948 (2020).

52. González-Ruiz, V., Fernández-Fernández, M. R. & Fernández, J. J. Structure-preserving Gaussian denoising of FIB-SEM volumes. *Ultramicroscopy* **246**, 113674 (2023).

53. Kremer, J. R., Mastronarde, D. N. & McIntosh, J. R. Computer visualization of three-dimensional image data using IMOD. *J Struct Biol* **116**, 71–76 (1996).

54. González-Ruiz, V., García-Ortiz, J. P., Fernández-Fernández, M. R. & Fernández, J. J. Optical flow driven interpolation for isotropic FIB-SEM reconstructions. *Computer Methods and Programs in Biomedicine* **221**, 106856 (2022).

55. Makovetsky, R., Piche, N. & Marsh, M. Dragonfly as a Platform for Easy Image-based Deep Learning Applications. *Microsc Microanal* **24**, 532–533 (2018).

56. Fernández, J.-J. & Li, S. An improved algorithm for anisotropic nonlinear diffusion for denoising cryo-tomograms. *Journal of Structural Biology* **144**, 152–161 (2003).

57. Moreno, J. J., Martínez-Sánchez, A., Martínez, J. A., Garzón, E. M. & Fernández, J. J. TomoEED: fast edge-enhancing denoising of tomographic volumes. *Bioinformatics* **34**, 3776–3778 (2018).

58. Martinez-Sanchez, A., Garcia, I., Asano, S., Lucic, V. & Fernandez, J.-J. Robust membrane detection based on tensor voting for electron tomography. *Journal of Structural Biology* **186**, 49–61 (2014).

59. Martinez-Sanchez, A., Garcia, I. & Fernandez, J.-J. A differential structure approach to membrane segmentation in electron tomography. *Journal of Structural Biology* **175**, 372–383 (2011).

60. Agulleiro, J.-I. & Fernandez, J.-J. Tomo3D 2.0--exploitation of advanced vector extensions (AVX) for 3D reconstruction. *J Struct Biol* **189**, 147–152 (2015).

61. Agulleiro, J. I. & Fernandez, J. J. Fast tomographic reconstruction on multicore computers. *Bioinformatics* **27**, 582–583 (2011).

62. Fernandez, J.-J. Computational methods for electron tomography. *Micron* **43**, 1010–1030 (2012).

63. Vallat, R. Pingouin: statistics in Python. *JOSS* **3**, 1026 (2018).

64. Waskom, M. seaborn: statistical data visualization. *JOSS* **6**, 3021 (2021).

# Figures

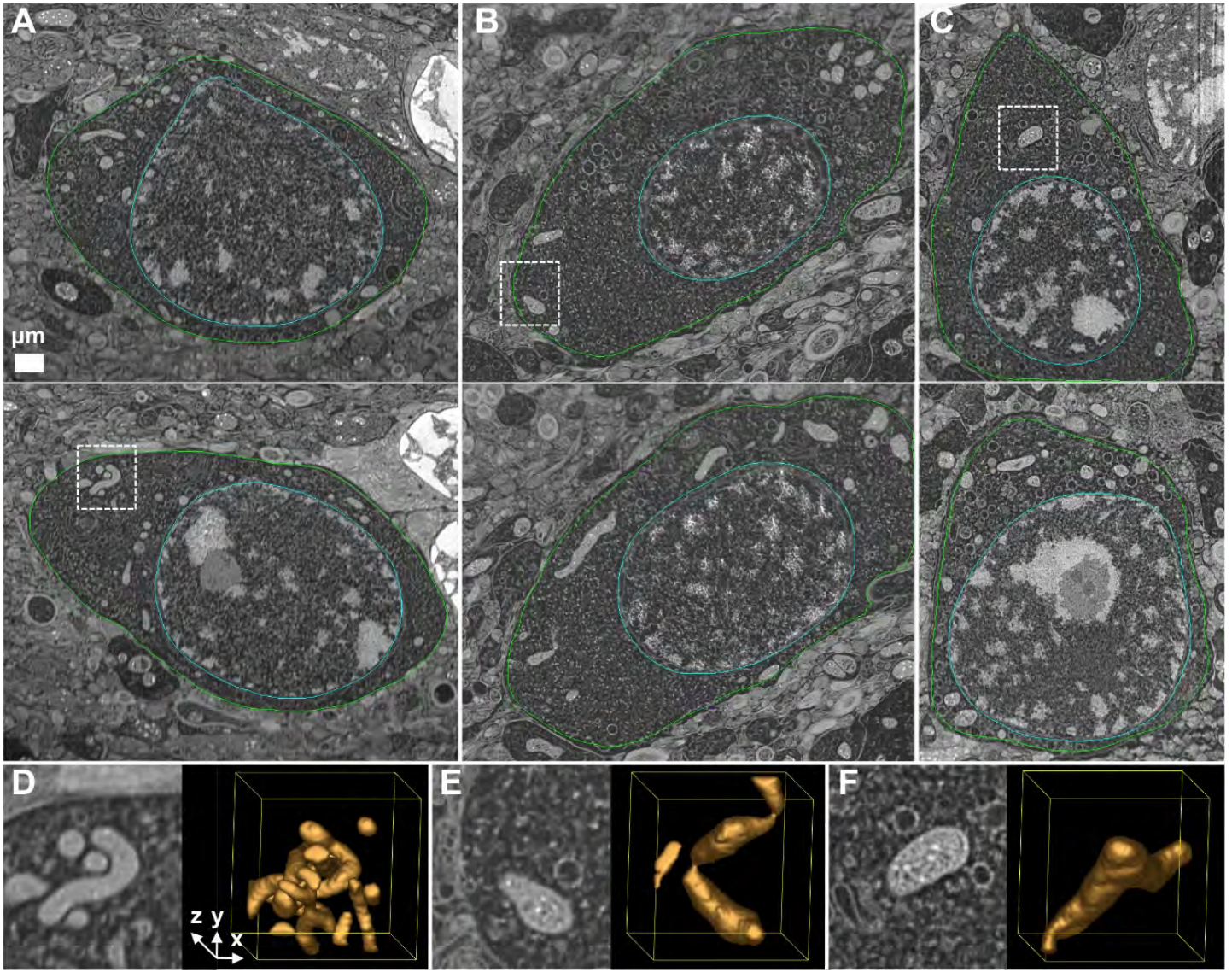

**Figure 1. FIB/SEM tomography of MSSNs.** (A,B,C) Two representative XY slices of one MSSN from the WT animal (A) and two MSSNs from the HD model (B,C) are shown. Green and cyan contours delineate the plasma and nuclear membranes, respectively. Mitochondria are identified as light grey masses inside the darker cytoplasm. Dashed boxes enclose selected cytoplasmic areas with representative mitochondria. Bar: 1 µm. **(**D,E,F) Magnified views of the dashed boxes in (A,B,C), respectively, are presented (left panels) along with 3D isosurface representations (right panels) of a volume of 2x2x2 µm$^3$ around those areas, thus showing the nearby context of those mitochondria.

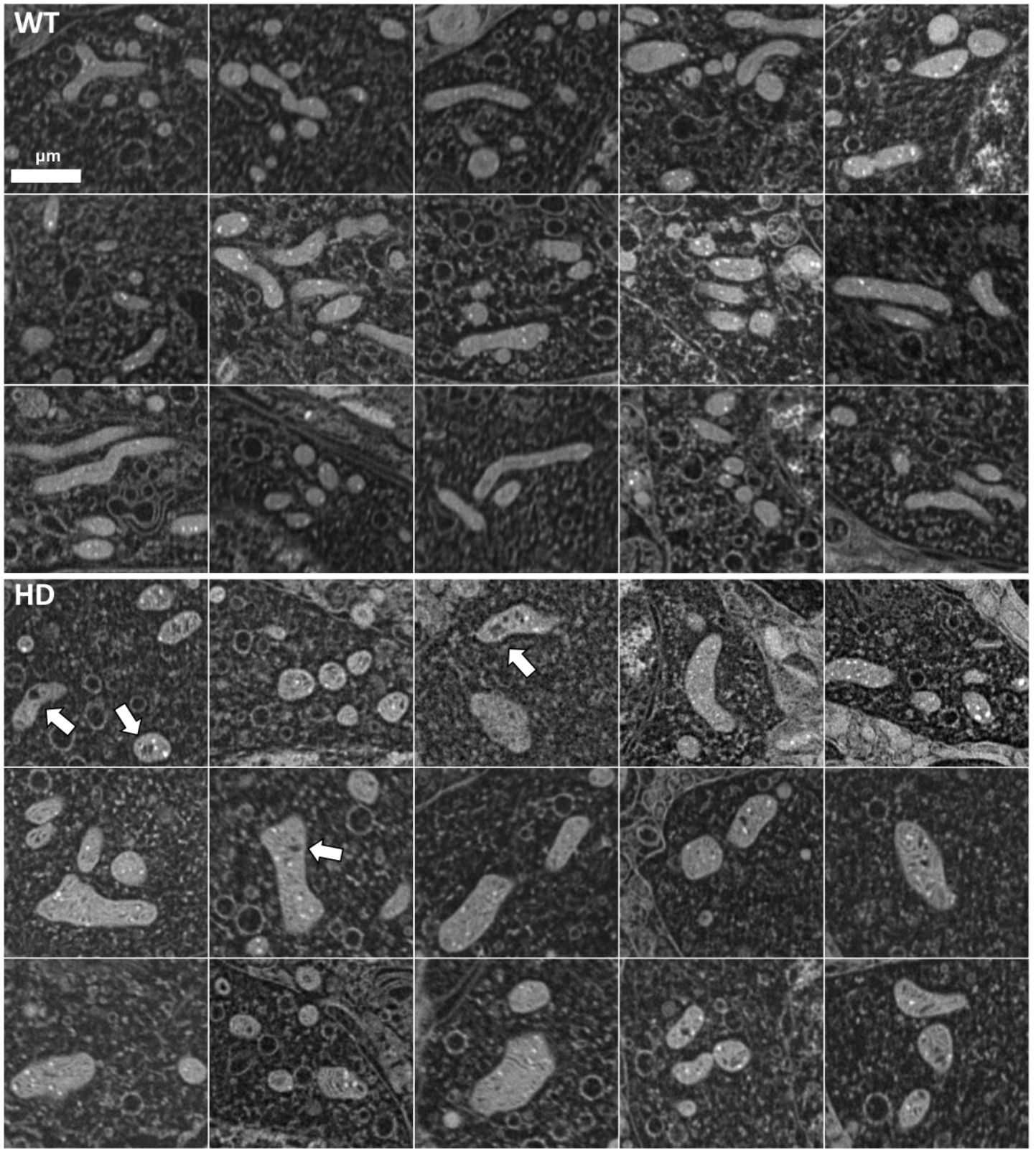

**Figure 2. Mitochondria in slices of FIB/SEM volumes.** Gallery of characteristic mitochondria observed in 2D slices of all FIB/SEM volumes. Top: mitochondria from 5 MSSNs from the WT animal. Bottom: mitochondria from 8 MSSNs from the HD animal model. Arrows indicate some hollow areas with fuzzy content in the matrix. Bar: 1 µm.

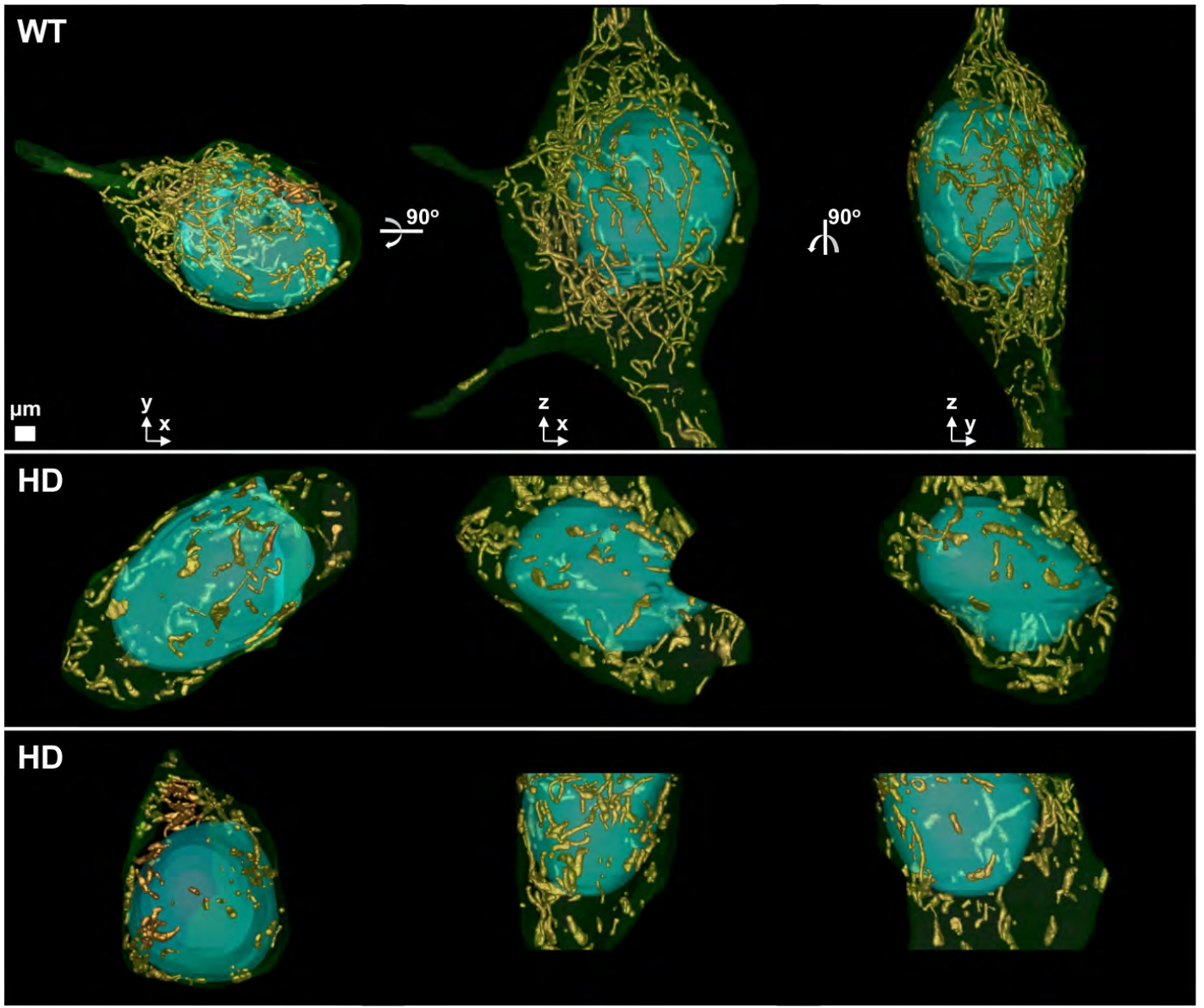

**Figure 3. 3D visualization of the FIB/SEM volumes**. Three different views of the volumes from the MSSNs in Figure 1 (A,B,C) are presented in the top (WT), middle (HD) and bottom (HD) rows, respectively. The leftmost views show the volumes with their Z axis running through the depth, a 90º rotation around the horizontal axis results in the views at the central panels, and a subsequent 90º rotation around the vertical axis produces the rightmost views. Segmented mitochondria are depicted with isosurface representation in gold color. Plasma and nuclear membranes are displayed in 85% transparent green and 50% transparent cyan, respectively, allowing visualization of the mitochondria behind the nucleus. The missing wedge in the volume shown in the middle row (central panel) is caused by a technical drift while FIB/SEM acquisition. Bar: 1 µm.

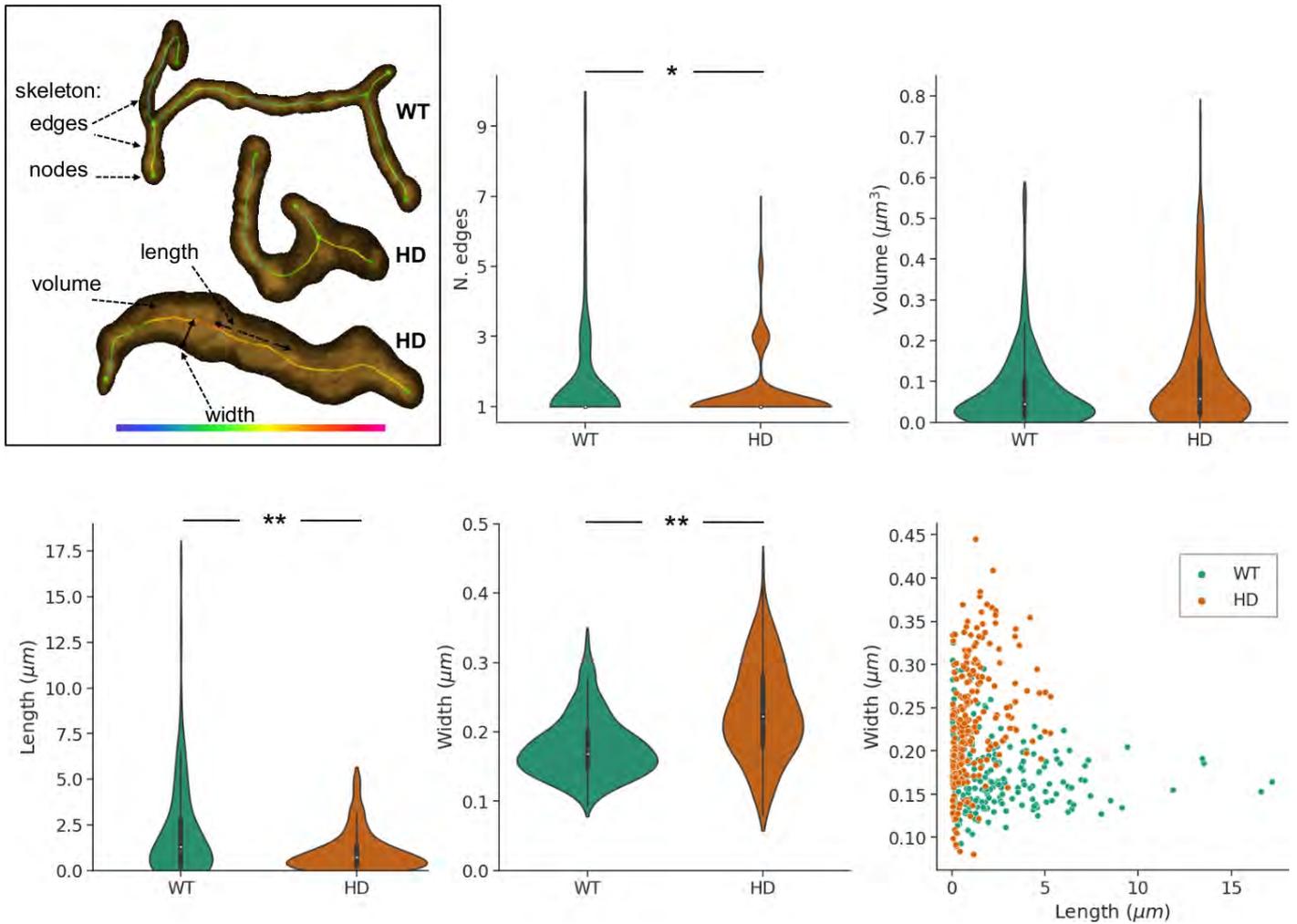

**Figure 4. Quantification of mitochondrial alterations in FIB/SEM volumes. Measurements** (upper left inset). Each individual mitochondrion consists of its body and skeleton, where the skeleton comprises edges (mitochondrial segments or branches) and nodes (i.e. terminal ends and branching points). For each mitochondrion, the following measurements are obtained: number of edges, volume, length (sum of the length of its edges) and width (average of the local width -here shown with a colormap- along its edges). Illustrative examples of mitochondria from WT and HD animals are presented in semitransparent 3D isosurface representation with their skeleton overlaid. **Quantification plots.** Comparison of measurements based upon 260 and 259 mitochondria from MSSNs of the WT animal and HD model. Violin plots show the distribution of the mitochondrial measurements. A miniature boxplot is included inside the violin plots, with the box representing the interquartile range (between the first and third quartile), an additional quartile with the whiskers and the median with a white dot. The scatterplot at the bottom rightmost panel represents measurements (length and width) of all individual mitochondria. *p<0.05; **p<0.0001.

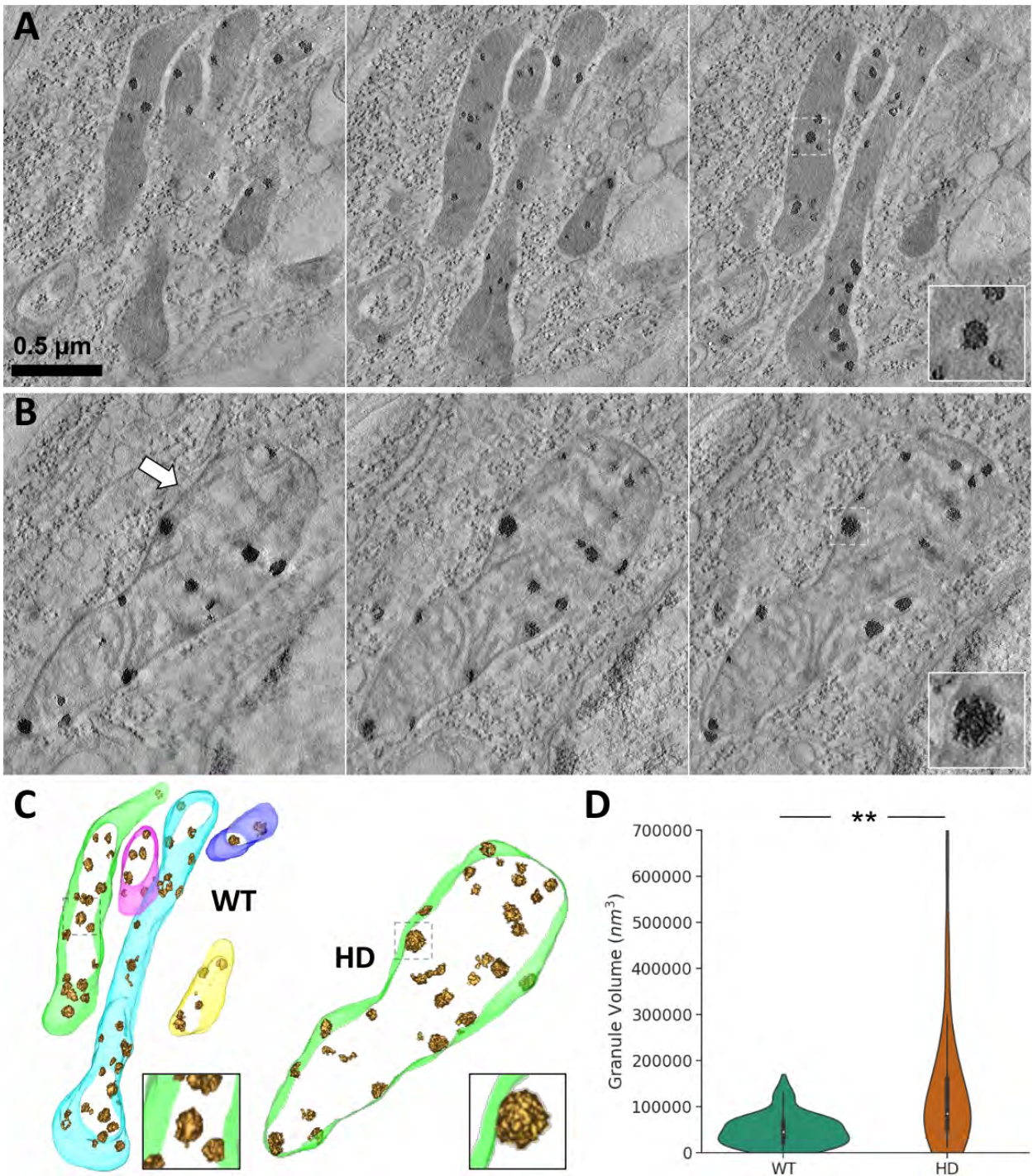

**Figure 5. Electron tomography of mitochondria from MSSNs and analysis of matrix granules.** (A,B) Tomograms of mitochondria from a WT animal (A) and a HD model (B). Three slices, separated by 22.12 nm, are shown for each tomogram. Note that the contrast is the opposite of that in FIB/SEM volumes (Figures 1 and 2), with electron-dense material appearing darker. The arrow points to a hollow area with fuzzy content. Dashed boxes indicate granules magnified in the insets. Bar: 0.5 µm. (C) 3D visualization of the matrix granules within mitochondria. Granules are presented with isosurface representation in gold color and mitochondrial membranes are delineated in different semitransparent colors. (D) The distribution of the granule volumes obtained from 66 and 37 granules of the WT and HD animals, respectively, are presented with violin plots (right). Similar to Figure 4, a miniature boxplot is included inside the violin plots and the median of the distributions denoted by a white dot. **$p<0.0001$.

31